\documentclass[11pt]{article}
\usepackage{fleqn,cospar}
\usepackage{url}
\usepackage{graphicx}
\usepackage[figuresright]{rotating}
\title{Models for the Origin of the Knee in the Cosmic-Ray Spectrum}

\author{A.D.Erlykin\address{P.N.Lebedev Physical Institute, 
 Leninsky pr. 53, Moscow 117924, Russia}$^{,2}$ and
A.W.Wolfendale\address{Physics Department,
 University of Durham, Durham DH1 3LE, UK}}

\begin{document}

\maketitle

\begin{abstract}
A sudden steepening of the cosmic-ray energy spectrum ( the knee ) is
observed at an energy of about 3 PeV (1 PeV = 10$^{15}$eV). The
experimental study of the PeV
cosmic rays has intensified greatly during the last 3 years. The recent
results on extensive air showers allow us to conclude that: a) the knee has
an astrophysical origin; b) the `sharpness' and the fine structure of
the knee rule out `Galactic Modulation' as the origin of the knee;
c) most likely the knee is the result of the explosion of a single,
recent, nearby supernova.
\end{abstract}

\section*{INTRODUCTION}

Cosmic rays spread over 11 decades of energy with an almost
featureless power law spectrum. There are just two structures which
are well established:
the steepening at an energy of about $3 \cdot 10^{6}$ GeV and the flattening
near $10^{10}$ GeV. The first is called {\em the knee}, the second is
{\em the ankle}. The first one was first found at Moscow University 42
years ago (Kulikov and Khristiansen, 1958), the second is
younger. Both features are crucial for understanding 
cosmic-ray originand propagation. We concentrate on the knee origin, because there
has been progress in its study during the last few years. This
progress is ensured by the work of new extensive air shower 
( EAS ) and Cherenkov arrays in Germany: KASCADE (Glasstetter {\em
et al.}, 1999), USA: CASA-MIA (Glassmacher {\em et al.}, 1998), DICE
(Swordy and Kieda, 2000), BLANCA (Fortson {\em et al.}, 1999), South
Pole: SPASE-2/VULCAN (Dickinson J.E. {\em et al.}, 1999), Canary
Islands: HEGRA (Arqueros {\em et al.}, 2000), Russia: TUNKA (Gress
{\em et al.}, 1999), Armenia: MAKET ANI ( Chilingarian {\em et al.},
1999) and others. These new experimental data let us make a
comprehensive analysis of the situation around the knee and this is a
subject of the present paper.

\section*{MODELS FOR THE ORIGIN OF THE KNEE}

Models of the knee proposed to date can be divided into two distinct
classes: {\em astrophysical} and {\em interaction} models. The 
{\em astrophysical}
models attribute the change in the spectra of the observed EAS
to the change in the energy spectra of the primary cosmic
rays. The {\em interaction} models imply that the primary energy spectrum
has no such sharp change and the observed steepening of EAS size
spectra is due to the sudden change of the nature of the interactions 
between the high energy particles of primary cosmic rays and the
atmosphere. The astrophysical models are more numerous and
developed. They might be also subdivided into two classes: the {\em source}
models, with a change of sources or their acceleration mechanisms, and
the {\em propagation} models, with a change of the cosmic-ray
propagation between the source and the observer. 

The very first model, which was proposed for the explanation of
the knee immediately after its discovery (Peters, 1959; Goryunov {\em
et al.}, 1962)
belongs to the latter class - it is the {\em Galactic Modulation} or
{\em diffusion} model which still has its supporters ({\em eg} Kalmykov
and Pavlov, 1999). The basic idea is that at low energies the
particle giroradius in the Galactic magnetic fields is small, their
motion between the magnetic irregularities is like a slow diffusion and
cosmic rays are trapped inside our Galaxy. However, Galactic magnetic
fields are not strong enough to trap high-energy particles which begin to
escape from the Galaxy: the higher the particle energy - the stronger
their leakage. This rising leakage results in the steepening of the
cosmic-ray energy spectrum.            

The {\em source} models associate the lack of high-energy particles
beyond the knee either with the increased loss of their energy in
intensive radiation fields (Hillas, 1979) or with the termination of
their acceleration mechanism (Biermann, 1993). The latter is often connected
with the ceasing of the shock wave produced by the supernova explosion. 

The {\em interaction} models explore the fact that we still observe
the knee indirectly, deep in the atmosphere, mostly by means of
EAS. They argue that the slope of the EAS size spectrum steepens because
interactions of high-energy particles suddenly change their character
at an energy of a few PeV ( Nikolsky, 1995 ).

\section*{EAS CHARACTERISTICS IN THE KNEE REGION}
The basic characteristics which are important for the analysis of
the knee origin are the primary energy spectrum, the mass composition of
the primary particles and the anisotropy of their arrival directions.
The first two determine the spectral shape and ratios between
different EAS components. 

\subsection*{Shape of the EAS size spectra}

Even the very first measurements of the EAS size 
spectrum revealed that the knee is surprisingly sharp. It would not
be so sharp if different primary nuclei have the knee at the same
rigidity, as in the Galactic Modulation model, and particularly if one
takes into account the huge fluctuations of the EAS size for the
showers of fixed primary energy observed in the lower part of the
atmosphere. However, it is sharp indeed. 

New measurements have raised the number of EAS size spectra to 40. There
are also 8 low energy muon size spectra, 3 high energy muon
multiplicity spectra, 3 hadron size and hadron energy spectra as well
as 5 spectra of the Cherenkov light emitted by the charged particles
of the shower. These measurements, in spite of the large spread of
their results, clearly show that:

$*$ the spectra of {\em all} components have a sharp knee at the values
corresponding to a primary energy of about 3 PeV. Traces of the
knee are seen even in high energy muon multiplicity spectra, in spite
of the tremendous fluctuations intrinsic for this component. Particularly
important is the observation of the knee in the spectra of Cherenkov
light, because these measurements give a good estimate of the total energy
of the cascade, and therefore of the primary.

This is an important result because it does not leave room for the 
interaction models.  
If all the components of the shower show steepening of their
spectra, then, in order to preserve the same slope of the primary
spectrum in the wide energy range, it is necessary to assume that primary
particles transfer their energy into some unobservable component. 
The only known unobservable component which remains undetectable up to date
is the neutrino. However, we know that the neutrino has only a weak interaction
and is born in weak decays. The energy fraction carried away by it
could not rise with the energy of the cascade, so that the neutrino cannot
be the component which takes away the missing energy of the cascade. 
The alternative hypothesis is the production of a
hypothetical heavy particle above a threshold energy $\sim$3 PeV with a large
and rising cross-section which escapes our detection due to exotic penetrating
properties. It is in contradiction with all our knowledge of multiparticle
production.

Such a conclusion does not exclude new features, which might
appear in particle interactions at high energies, particularly in
nucleus-nucleus collisions. It states only that such new features, if
they exist, could not be the origin of the knee.

$*$ the EAS size corresponding to the knee position decreases
with the atmospheric depth at which the size is measured, in
accordance with expectation, if the knee occurs at a fixed primary
energy ( Figure 1 ).    
\begin{figure}[htb]
\begin{center}
\includegraphics[height=75mm,width=14cm]{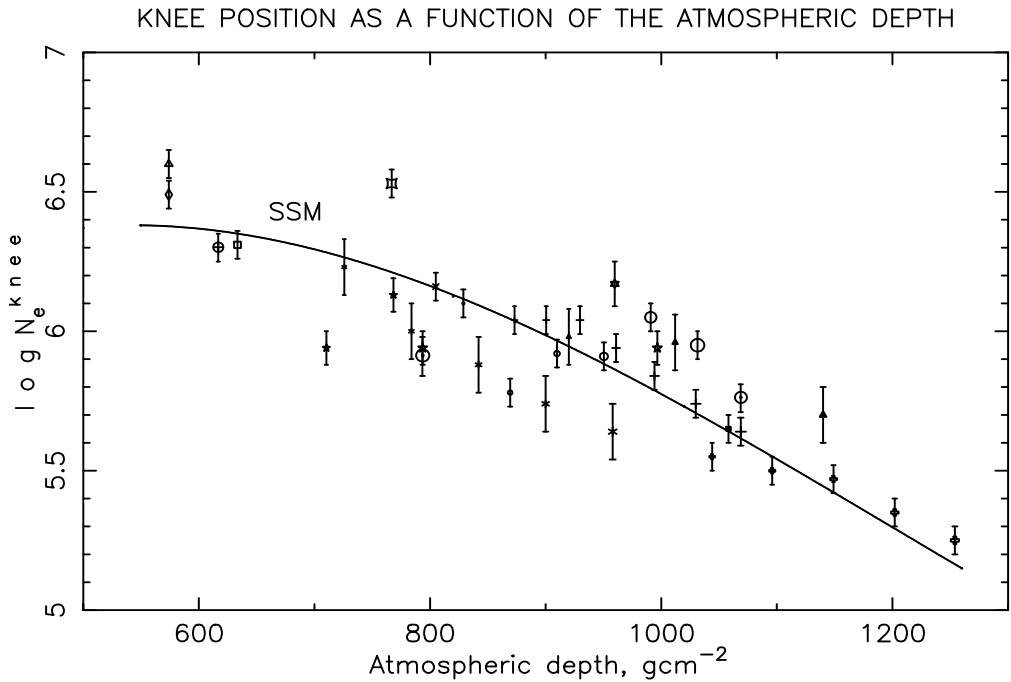}
\end{center}
\caption{The EAS size corresponding to the knee position as a function of the atmospheric depth at which the size is measured. The full
line denoted  as 'SSM' is calculated from our Single Source Model.}
\label{fig:model1}
\end{figure}

$*$ The sharpness of the knee exceeds that expected in the Galactic Modulation
model.  If one determines the sharpness as 
 $S = -\frac{d^2(logI)}{d(logN)^2}$, where $I$ and $N$ are the EAS
intensity and size correspondingly, then the expected 
sharpness in the 
Galactic Modulation model should not exceed 0.3 even for the primary
spectrum itself. The real situation is shown in Figure 2. It is seen that 
{\em all} measured values are above 0.3. The overall mean is about 
1.3$\pm$0.1
for the EAS size spectra and for Cherenkov light spectra it is even
higher - around 3 ( the reason for the higher value is that it relates
more nearly to the primary spectrum ).\\ 
\begin{figure}[htb]
\begin{center}
\includegraphics[height=75mm,width=14cm]{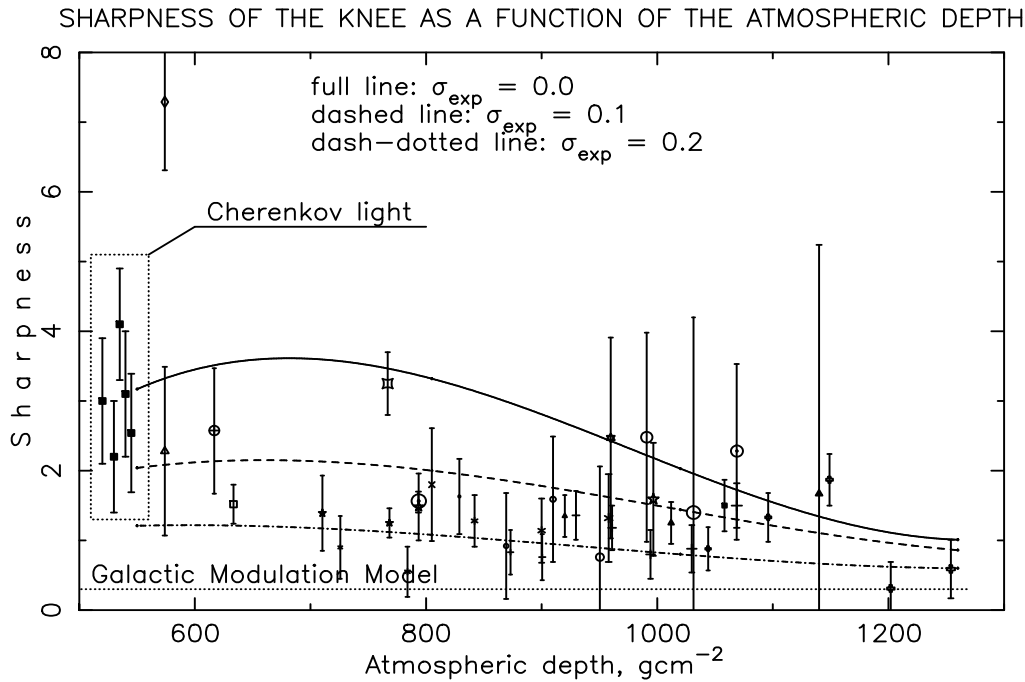}
\end{center}
\caption{Sharpness of the knee from all available EAS experiments as a
function of atmospheric depth. The lines are the expectation from our
SS model for different values of experimental errors: $\sigma_{logN_e} = 0$
- full line, 0.1 - dashed line and 0.2 - dash-dotted line. Dotted line
- Galactic Modulation model. Also shown are Cherenkov results 
(~inside dashed box~).}
\label{fig:model2}
\end{figure}
$*$ The increasing number of observations has given birth to another 
interesting finding: the fine structure of the knee. In order to study
the shape of the spectra around the knee when the positions of the
knee in the individual spectra spread over a decade ( Figure 1 ), the
EAS sizes and energies should be normalized to those 
corresponding to individual knee positions. The knee position was
determined 
as the point of the maximum sharpness of the spectrum. If all the spectra are
normalized at their knee point, 
then, at the size which is by the factor of 4 higher than
the size at the knee, there is another intensity peak with a positive 
sharpness 
(Erlykin and Wolfendale, 1997). It is not so distinct as the knee,
because, due to the larger statistical and systematic errors of each
individual experiment at high energies, it is often hidden within the
error bars. However, superimposing 
individual spectra with a normalization at the knee unveils it
unambiguously. The superposition in this approach is similar to the method in
which any weak, but regular, signal is searched for beneath a
stochastic background. 

The first evidence of this peak was found on the basis of just 9
EAS size spectra. The increasing number of measurements has let us confirm
its presence with much higher confidence. Moreover, the second peak is
found also in Cherenkov light spectra, which proves its astrophysical origin. 
These results are shown in Figure 3. Here, the fine structure of the 40 EAS
size and 5 Cherenkov light spectra are shown not for the
sharpness, but for the excess of the intensity over the running mean 
(Erlykin and Wolfendale, 2000).
The running mean was determined in the range of $log(Size)$ equal to
$\pm 0.25$ around each measured point. The presence of the second peak
in the excess is seen beyond the error bars both in EAS size and in
Cherenkov light spectra.
\begin{figure}[htb]
\begin{center}
\includegraphics[height=12cm,width=15cm]{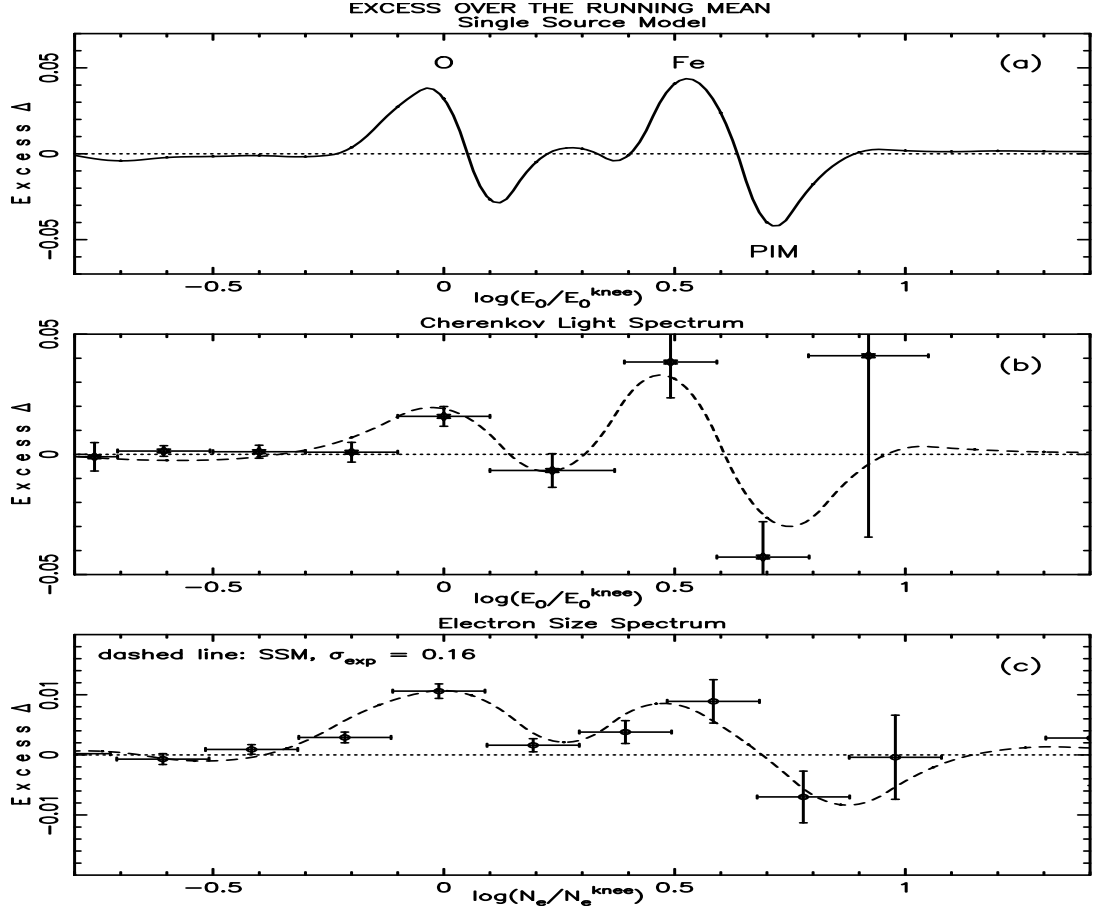}
\end{center}
\caption{Excess over the running mean. The results relate to the
average excess, in $\Delta log(E_0/E_0^{knee})$ and $\Delta
log(N_e/N_e^{knee})$ bins equal to 0.2, for all the world's data. The topmost
curve is from the single source model, PIM denotes 'post iron
minimum'. The middle graph is for the Cherenkov results and the lowest for the
electron size spectrum. The dashed curves for the Cherenkov and electron
spectrum graph are the straightforward SSM predictions; that for
Electron Size Spectrum includes an `experimental error' with standard
deviation $\sigma_{exp} = 0.16$ (~in logarithmic units~).}
\label{fig:model3}
\end{figure}
These last results on the magnitude of sharpness and the existence of the
second intensity peak are also important for the problem of the origin
of the knee, because they leave
no room for the Galactic Modulation model with its smooth and regular
steepening of all the constituent nuclei spectra.  

\subsection*{Mass composition}   

The primary mass composition in the knee region is still a matter of
hot debate. The problem is that direct measurements in space do not
yet reach the important PeV region. All the studies of the mass
composition there are indirect and based on the ratios between different
shower components. The range of conclusions is highly disparate;
however there has been progress here in the last few years. It has 
become possible because more EAS arrays detect and analyse not just
one but several shower components simultaneously. 

Most of the experiments now give convergent results and conclude
that the mass composition becomes heavier beyond the knee. The
difference is confined to the estimate of the mass composition below the knee
and at the knee itself. The problem is that the primary mass estimated
on the basis of indirect measurements often depends on the
shower components chosen for the analysis. For instance,
the primary mass, derived from the electron and muon components, is
lighter than that derived from electron and hadron components (Kampert 
{\em et al.}, 1999). The analysis is always made using most
reliable interaction models. Does this inconsistency indicate some
drawbacks in our understanding of high-energy interactions ? This is
not unlikely, particularly as far as the nucleus-nucleus interactions is
concerned.
 
Two conclusions might be drawn on this point:\\
$*$ the methods used for the analysis of the mass composition at high
energies should be tested and calibrated at lower energies, where
direct measurements exist and where we are sure about our interaction
models;\\
$*$ the estimates of the primary mass and the most reliable
interaction model should be based on the analysis of the
maximum number of shower components. 

\subsection*{Anisotropy} 

The anisotropy of the arrival directions can provide valuable
information about the origin of the knee. The overall situation has
been discussed by us in Erlykin, Lipsky and Wolfendale, 1998 and is shown
in Figure 4, so that we shall not go into details here.  We only
underline that both the amplitude and the 
phase of the first anisotropy harmonic show sharp changes in the
PeV-region. It is another argument in favour of an astrophysical
origin and against the interaction model of the knee.  
\begin{figure}[htb]
\begin{center}
\includegraphics[height=12cm,width=15cm]{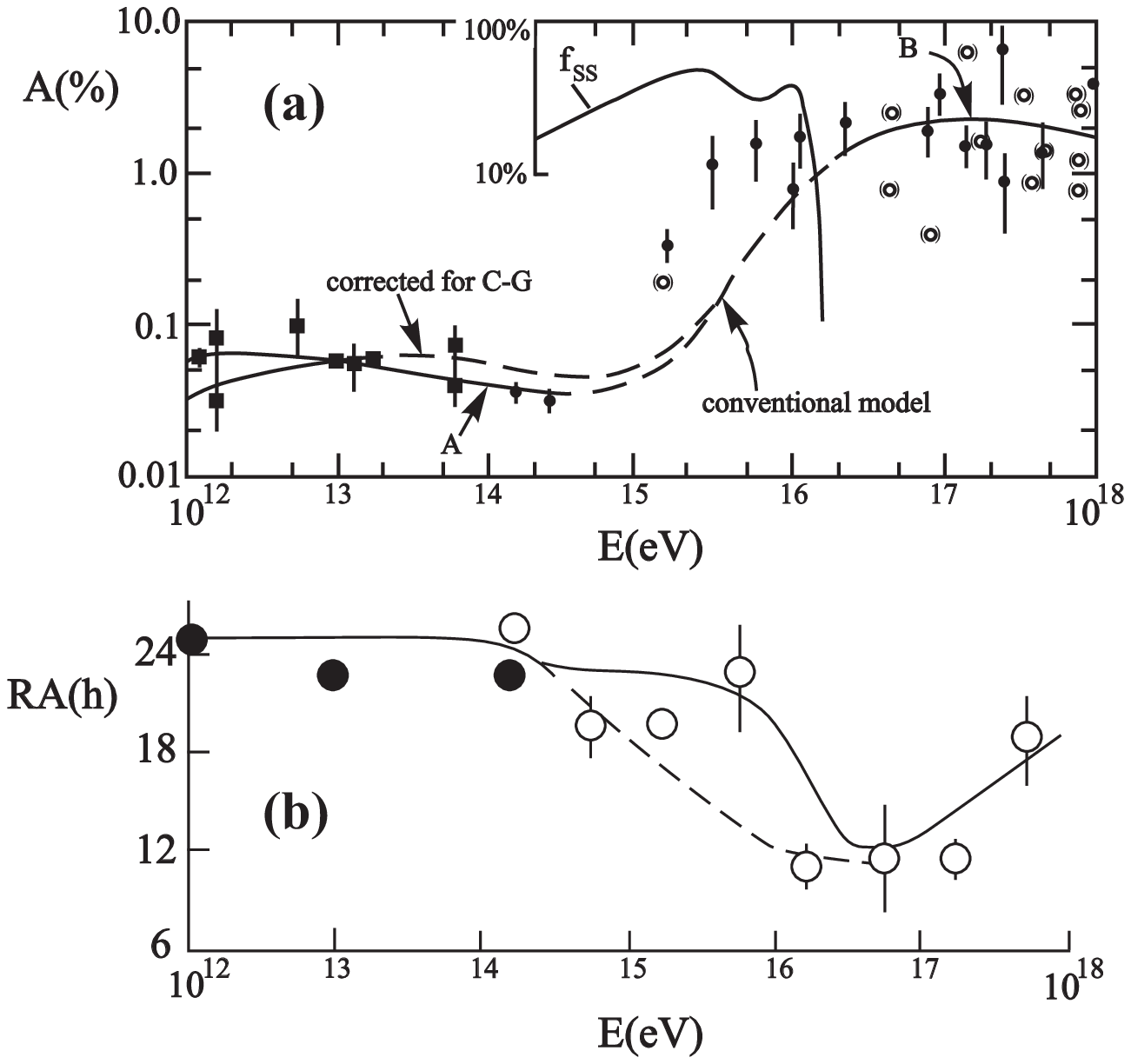}
\end{center}
\caption{Anisotropy amplitude (a) and phase (b) (first harmonic) from
Erlykin, Lipsky and Wolfendale, 1998. (a) Filled symbols represent the
experimental data, the bracketed circles are of low statistical
precision. The full lines A, B and dashed line in the range
$10^{15}-10^{16}$ eV are our 'conventional model' estimate. A tentative
correction to the lowest anisotropy amplitudes has been made for the
Compton-Getting Effect. The inset shows the fraction of the total flux
derived from the Single Source ( denoted f$_{ss}$ ). (b) Open circles
represent the experimental data, the filled circles are after the tentative
correction for the Compton-Getting Effect. The dashed line is the
prediction of the conventional model, the full line is for our SSM.}
\label{fig:model4}
\end{figure}
\section*{SINGLE SOURCE MODEL OF THE KNEE (SSM)}
There is a general conjecture, based on the energy and theoretical
arguments, that supernova explosions are responsible for the formation
of the cosmic ray energy spectrum below, and possibly even beyond, the
knee. The intensity of their explosions is correlated with the star   
forming regions, whose properties: density and temperature of the
interstellar gas, strength and irregularity of magnetic fields {\em
etc} vary over a wide range. Thus, any kind of averaging over the
range of supernovae would eventually result in a smoothly varying
cosmic-ray spectrum. In 1997 we put forward a model in which the knee
is formed by the explosion of just {\em a single, nearby and recent
supernova}. The spectrum of cosmic rays from the shock caused by that explosion protrudes
through the smooth background formed by many other ( unspecified ) 
sources ( Figure 5 ).    

\begin{figure}[htb]
\begin{center}
\includegraphics[height=12cm,width=15cm]{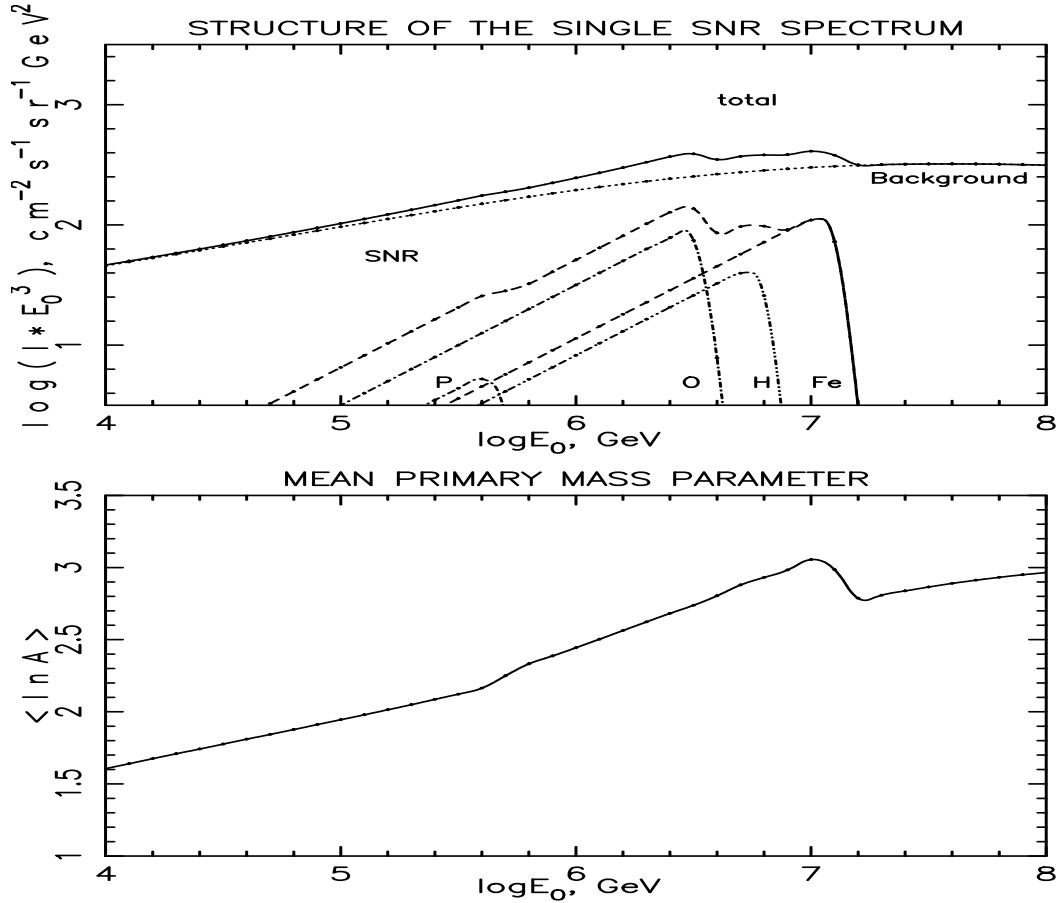}
\end{center}
\caption{The Single Source Model of the primary energy spectrum.
(a) Structure of the spectrum. SNR denotes the contribution from the
single supernova and lines denoted as P, O, H, Fe - the contribution of the constituent spectra from protons, oxygen, heavy ( Ne-S ) and iron nuclei correspondingly. The background spectrum is assumed to be due to many 
(unspecified) sources. (b) Mean mass parameter in SSM.}
\label{fig:model5}
\end{figure}
The shape of the SSM energy spectrum has been taken from the theory of
supernova remnant acceleration (~Berezhko {\em et al.}, 1996~). The specific
features of that theory are a flat energy spectrum $\sim E^{-2}$ in
the shocked region, a quite high maximum energy of accelerated
particles corresponding to the rigidity about  $R_{max} \approx$ 0.4
PV and a very sharp cutoff of the spectrum beyond that rigidity. If the
explosion occured in the hot interstellar medium, which is the case of
the Local Bubble, in which our solar system is immersed, then different
nuclei should be well separated in the energy scale, because the
energies of the accelerated nuclei are proportional to their charge $Z$.
We attribute the knee to the contribution of oxygen nuclei,
because:\\
(i) its position at 3 PeV corresponds to the theoretical prediction of
$E_{max} \approx R_{max}Z$ for oxygen accelerated in the hot
interstellar medium;\\
(ii) its position and intensity stand well at the extrapolation from the
results of direct measurements at lower energies;\\
(iii) it helps to understand the sharpness of the EAS size spectra,
because the nuclei-initiated showers have smaller fluctuations in
their development and the sharp cutoff in the primary energy spectrum
should not be diluted when transferred to the EAS size spectrum.      
\subsection*{Sharpness}
The assumptions adopted in SSM help us to understand the observed
sharpness of all the shower components. It is ensured by the fact that
it is just one single supernova, exploded in a single environment,
it is only one nucleus - oxygen, dominating at the knee energy and oxygen
is heavy enough to minimize fluctuations of the shower development. 

In Figure 2 there are three lines which show the expected behaviour of
the sharpness as a function of the atmospheric depth in SSM, for
three values of the experimental errors in the determination of the
shower size $N_e$. The best fit value of standard deviation
$\sigma_{logN_e}$ is 0.16, which is quite reasonable for most of
the experimental EAS arrays.

The model gives also a satisfactory description of the smooth
decrease of the sharpness with atmospheric depth.
\subsection*{Mass composition}
(i) If the knee is attributed to oxygen then the second peak is
probably 
associated with iron. The fit of the experimental data requires some
medium heavy nuclei of Ne-S group to be also present. This assumption
explaines well the existence of the second peak and its separation
from the knee by a factor proportional to the charge (~see SSM
based curves in Figure 3~);\\ 
(ii) at the moment it is not possible to rule out the alternative
association of the first knee peak with helium, and the second with
oxygen ( or CNO group ). However, this alternative is less
favorable because the larger fluctuations in the helium-induced showers
lead to a stronger dilution of the sharpness in the EAS size
spectra. The smooth
decrease of the sharpness with atmospheric depth requires also the
mass composition of the background to be lighter than that of the
single source, thus in the case of helium dominating at the knee the
background should consist mostly of protons, which is unlikely.\\
(iii) in both versions of the SSM the primary mass should rise with
energy beyond the knee, which agrees with the results of the
experiments discussed in subsection 3.2. In Figure 5b this rise is
shown by a smooth line. It is seen that in the SSM the rise is
continuous, which agrees with the tendency found at lower energies.
\subsection*{Anisotropy}
Despite the fact that the source in the SSM is recent and nearby it is certain
that it should not create a very strong anisotropy. If our
interpretation of the mass composition, i.e. the dominance of oxygen
and iron in the peaks is correct, then the peak energies correspond 
to a rigidity of 0.4 PV. The maximum Larmor radius of all
the nuclei at a rigidity of 0.4 PV in the surroundings of the solar
system is about 0.1 pc. The propagation of the cosmic rays from the
source is definitely by diffusion and the anisotropy is therefore determined
just by the gradient of the very local cosmic-ray density. The second factor
which might be important is the location of the solar system with
respect to the shock front. If we are inside it, the cosmic rays are
highly isotropized and even their gradient is not easy to
detect. Perhaps the change of the amplitude and the phase of the first
harmonic seen in Figure 4 are the only imprint of the nearby source on
the generally isotropic flux of the cosmic rays at PeV energies ? 
\section*{On the way to the identification of the single source}
There are not many of single sources which could be classified as 
`nearby and recent' {\em ie} within the range of a few hundred parsecs
and a few hundred thousand years ( Figure 6 ). 

 The first step in the identification of the
source responsible for the knee is to determine whether we are inside
or outside the shock front. If our single source is indeed a
supernova, the theory of cosmic-ray acceleration by its shock wave 
indicates that outside the shock region only the highest-energy cosmic
rays reach the observer. This spectrum is very hard and looks like a
line near the maximum energy (Berezhko {\em et al}, 1996). We tested
this type of the spectrum by applying it to our model and trying to fit
the observed experimental data on the sharpness  and the behaviour of
excess over the running mean. The sharpness of the knee in the primary
energy spectrum obtained in this model grew up to 6.0, the excess over
the running mean in the knee increased by a factor of 2 compared
with that indicated in Figure 3 and a large region of negative excess
appeared below the knee both in the Cherenkov light and
in the EAS size spectra, all these features are completely inconsistent with what
is observed in the experimental data. Therefore, on the basis of this
analysis, we should say that the case when we are {\em inside the shock} is
preferable, compared with the opposite case when we are outside it.
We can remark that this conclusion helps to understand also the relatively
small amplitude of the anisotropy in the knee region. The typical
propagation of the shock wave from the supernova explosion, taken from 
the Berezhko {\em et al.}, 1996 calculations, is shown in Figure 6 by the
dotted line. The sources inside the shock should lie to the right of
this line.
\begin{figure}[htb]
\begin{center}
\includegraphics[height=9cm,width=14cm]{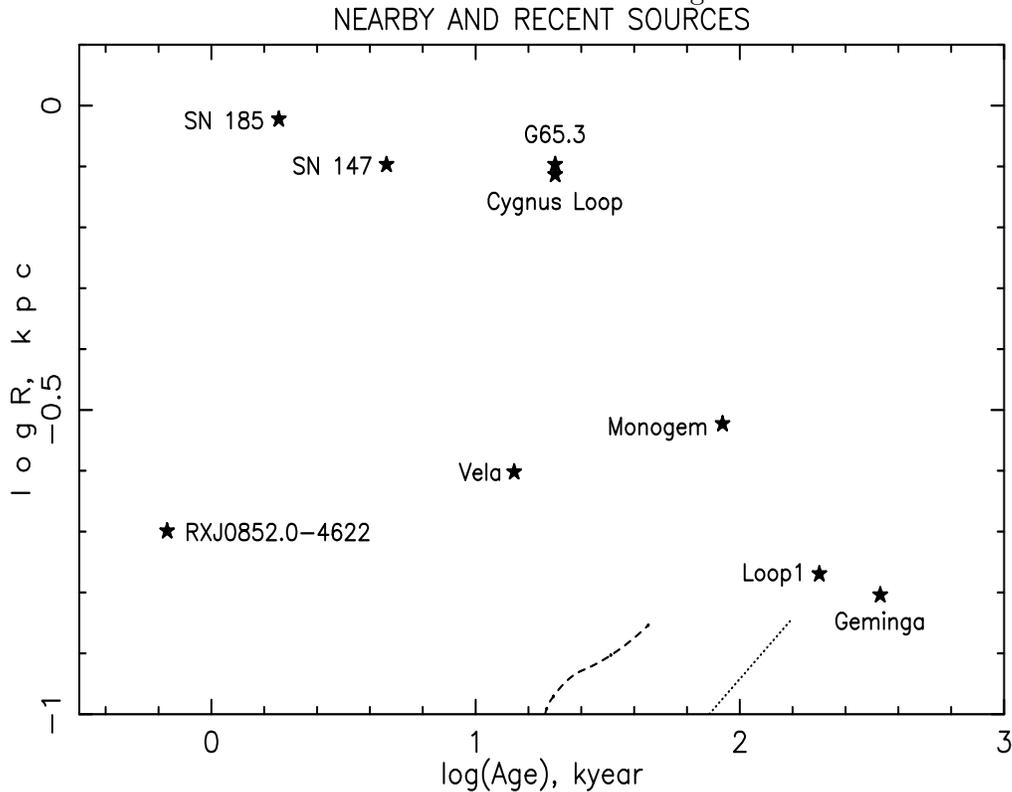}
\end{center}
\caption{Plot of recent and nearby sources, which might be responsible
for the formation of the knee. The dotted line shows the typical
propagation of the shock front. The dashed line is derived from the
comparison of the energy contained in cosmic rays for our Single Source
and the energy content of the cosmic rays accelerated by the supernova
remnant in the calculations of Berezhko {\em et al.}, 1996 calculations. The possible source
candidates should lie to the right of these lines.}
\label{fig:model6}
\end{figure}

Another step in the identification of the source might be based on the
analysis of the energy content of the source. The candidate
for the source could not be too far from the solar system or too close
to the moment of the explosion in order to have enough energy in
cosmic rays and give the required contribution to the cosmic-ray flux
at the knee. We have compared the energy density of the our single source, 
shown in Figure 5a, with that contained in the spectra calculated by
Berezhko {\em et al.}, 1996. The result is shown by the dashed line in
Figure 6 and means that required energy density might be achieved if the
supernova is to the right of this line. Thus the possible candidates
for our single source might be found in the right and lower corner of
the plot in Figure 6. At the moment, both the source which gave birth to the
Loop I and Geminga pulsar are the most favorable contenders. However,
taking into account large
errors in the determination of the age and distance, other nearby
sources, shown in the Figure 6 \underline{or still unknown}, cannot
be excluded. The problem of the identification is still with us.
\section*{CONCLUSION}  
The recent results on extensive air showers (~both by way of particles
and Chherenkov radiation~) allow us to conclude that: a)
the knee observed in the cosmic-ray spectrum at about 3 PeV
has an astrophysical origin; b) the sharpness and the fine structure of
the knee rule out the Galactic Modulation Model as the origin of the knee;
c) the most likely model of the knee origin is the {\em source} model in
which the knee appears as the result of the explosion of a single,
recent, nearby supernova.       
\section*{Acknowledgments}
A.D.E. thanks the Organizing Committee of 33d COSPAR Assembly for the 
invitation to give this talk and for providing the financial support. 
The Royal Society and UK's PPARC are thanked for supporting this work.

\end{document}